\documentclass[journal]{IEEEtran}

\ifCLASSINFOpdf
\else
 
\fi

\usepackage{graphicx}
\usepackage{float}
\usepackage{amsmath}

\begin{document}

\title{Effects of Memristors on Fully Differential Transimpedance Amplifier Performance}

\author{Berik Argimbayev, Olga Krestinskaya and Alex Pappachen James\\
 Department of Electrical and Computer Engineering\\
Nazarbayev University,
Astana, Kazakhstan}

\maketitle


\IEEEpeerreviewmaketitle

\begin{abstract}
The progress of the Internet of Things(IoT) technologies and applications requires the efficient low power circuits and architectures to maintain and improve the performance of the increasingly growing data processing systems. Memristive circuits and substitution of energy-consuming devices with memristors is a promising solution to reduce on-chip area and power dissipation of the architectures. 
In this paper, we proposed a CMOS-memristive fully differential transimpedance amplifier and assess the impact of memristors on the amplifier performance. The fully differential amplifiers were simulated using $180nm$ CMOS technology and have 5.3-23MHz bandwidths and 2.3-5.7k$\Omega$ transimpedance gains with a 1pF load. We compare the memristor based amplifier with conventional architecture. The gain, frequency response, linear range, power consumption, area, total harmonic distortion and performance variations with temperature are reported.

\end{abstract}


\section{Introduction}
Communication between various devices is the principal idea of the IoT \cite{krestinskaya2017hierarchical,ISCASs} and the progress of transceivers plays a major role in the IoT development \cite{krestinskaya2017hierarchical,Ghorbani,Mumtaz}. Amplifier is a fundamental element in any analog circuit for IoT applications, and a majority of transceivers include transimpedance amplifiers (TIA) \cite{Ghorbani,Garcia,Saso,Masuch,Selvakumar}.
IoT systems require low power devices and circuits with small on-chip area, and ability to tolerate temperature variations \cite{krestinskaya2017hierarchical,Chang,irmanova2018neuron}. There are several research works attempting to fulfill these requirements in TIAs \cite{Djekic,Abd,Taghavi,Mekky}. However, the energy efficient amplifiers that can be used for large scale IoT applications is still an open problem \cite{ISCASs}. 

In this paper, we proposed the energy efficient memristor-based TIA design that can be used for large scale IoT architectures.
The use of memristors in various architectures have proven the to be efficient for the reduction of on-chip area and power consumption, comparing to the conventional CMOS-based designs \cite{krestinskaya2017hierarchical,ISCASs,irmanova2018neuron,dastanova}. This paper compares the conventional TIA design shown in \cite{Garcia} and proposed CMOS-memristive modification, where one resistor, two CMOS transistors are replaced by memristors. 

The rest of the paper is organized as follows. In Section II, mathematical analysis is conducted to find the expected value of the transimpedance gain in first and second designs. In Section III, gains, frequency responses, linear ranges, power consumptions, total harmonic distortions and performance variations with temperature are reported for all four designs. Section IV discusses differences between the presented designs and Section V concludes the paper.

\section{Proposed design and gain calculation}

\begin{figure*}[ht]
\centering
\includegraphics[width=\textwidth]{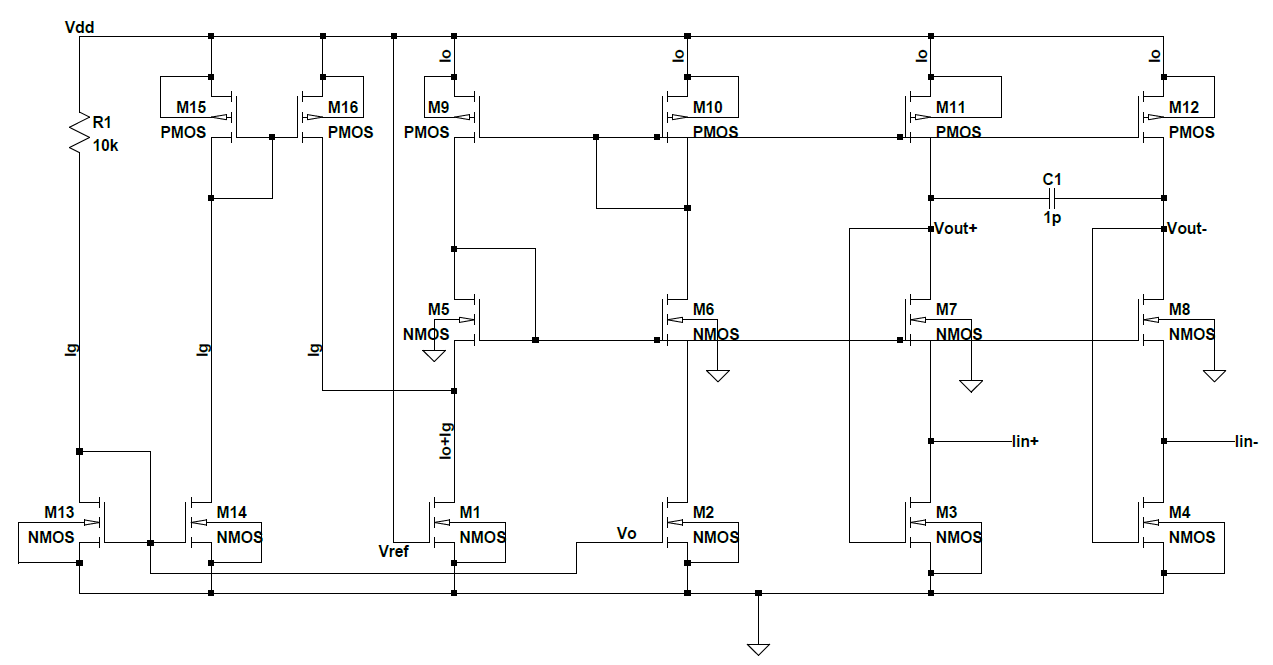}
\caption{Schematic of the differential amplifier without memristors \cite{Garcia}.}
\label{schemtran}
\end{figure*}

The schematics of fully differential TIA is presented in Fig. \ref{schemtran} \cite{Garcia}. We propose to replace resistor R1, and transistors M6 and M9 with memristors, which allows to reduce the on-chip area and power dissipation, and does not deteriorate the functionality of the circuit. 

In this amplifier, current $I_G$ is represented by Eq. \ref{e1}, where $r_{ds13}$ is the drain to source resistance of transistor M13 and $V_O$, which is the gate voltage of M2, equals to $V_{dd}-I_GR_1$.
\begin{equation}
\label{e1}
I_G = \frac{V_{dd}}{R_1+r_{ds13}}
\end{equation}
The drain current of M2 is represented by $I_O$, which is replicated by current mirror built by M9-M12.
The two other current mirrors, M13-M14 and M15-M16, replicate $I_G$, which is added to $I_O$ and is a drain current of M1. In case of M3 and M4, input currents $I_{in+}$ and $I_{in-}$ are added to $I_O$ respectively.

For correct operation transistors M1-M4 operate in active region, shown by Eq.2-5, where transistors M1-M4 have the same geometrical parameters and are all NMOS transistors, and $V_{ds}$ is the same to ensure the ideal matching condition.
\begin{align}
I_{D1}=I_O+I_G = \mu_NC_{ox}\frac{W}{L}V_{ds}(V_{dd}-V_T-\frac{n_N}{2}V_{ds})\\
I_{D2}=I_O = \mu_NC_{ox}\frac{W}{L}V_{ds}(V_{O}-V_T-\frac{n_N}{2}V_{ds})\\
I_{D3}=I_O+I_{in+} = \mu_NC_{ox}\frac{W}{L}V_{ds}(V_{out+}-V_T-\frac{n_N}{2}V_{ds})\\
I_{D4}=I_O+I_{in-} = \mu_NC_{ox}\frac{W}{L}V_{ds}(V_{out-}-V_T-\frac{n_N}{2}V_{ds})
\end{align}
Equation \ref{e2} is a result of subtracting Eq.3 from Eq.4.
\begin{equation}
\label{e2}
I_{in+} = \mu_NC_{ox}\frac{W}{L}V_{ds}(V_{out+}-V_O)
\end{equation}
Equation \ref{e3} is obtained by subtracting Eq.3 from Eq.2.
\begin{equation}
\label{e3}
I_{G} = \mu_NC_{ox}\frac{W}{L}V_{ds}(V_{dd}-V_O)
\end{equation}
Equation \ref{e4} is derived by dividing Eq.6 by Eq.7, and similar for Eq.\ref{e5}.
\begin{align}
\label{e4}
\frac{I_{in+}}{I_G}=\frac{V_{out+}-V_O}{V_{dd}-V_O}
\end{align}
\begin{align}
\label{e5}
\frac{I_{in-}}{I_G}=\frac{V_{out-}-V_O}{V_{dd}-V_O}
\end{align}
Eq. \ref{e6} can be obtained by rearranging Eq.8-9, where $\frac{V_{dd}-V_O}{I_G}$ and $R_1$ is determined by Eq.\ref{e7}.
\begin{equation}
\label{e6}
V_{out+}-V_{out-}=\frac{V_{dd}-V_O}{I_G}(I_{in+}-I_{in-})
\end{equation}
\begin{equation}
\label{e7}
R_1=\frac{V_{out+}-V_{out-}}{I_{in+}-I_{in-}}=\frac{V_{out}}{I_{in}}
\end{equation}

In our design, we replace R1, M6 and M9 by memristors, and the analysis from Eq.11 is useful to determine $R_{off}$ (Eq.12). 
\begin{equation}
R_{off}=R_1=\frac{V_{out+}-V_{out-}}{I_{in+}-I_{in-}}=\frac{V_{out}}{I_{in}}
\end{equation}
As memristors are programmed to have OFF state resistance $R_{off}$ equal to 10k$\Omega$, the substitution of M6 and M9 does not have effect on the above calculations.

\section{Simulation results}
\begin{figure}[!t]
\includegraphics[width=\columnwidth]{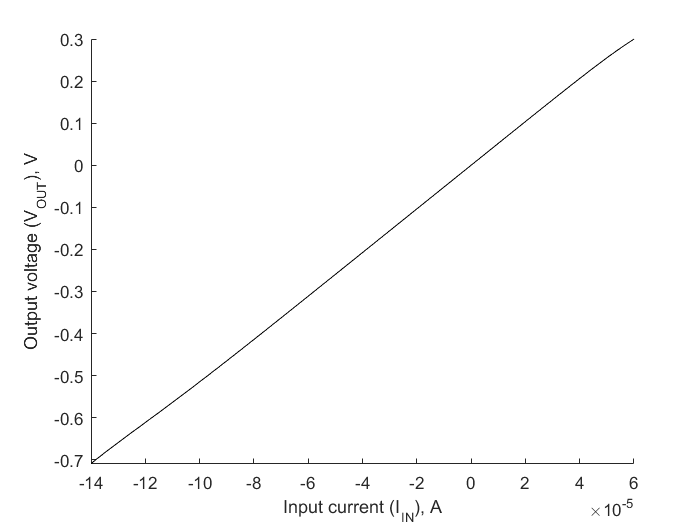}
\caption{Output voltage V$_{OUT}$ versus input current I$_{IN}$. Original geometric parameters, transistor only case}
\label{tranl1gain}
\end{figure}

\begin{figure}[!t]
\includegraphics[width=\columnwidth]{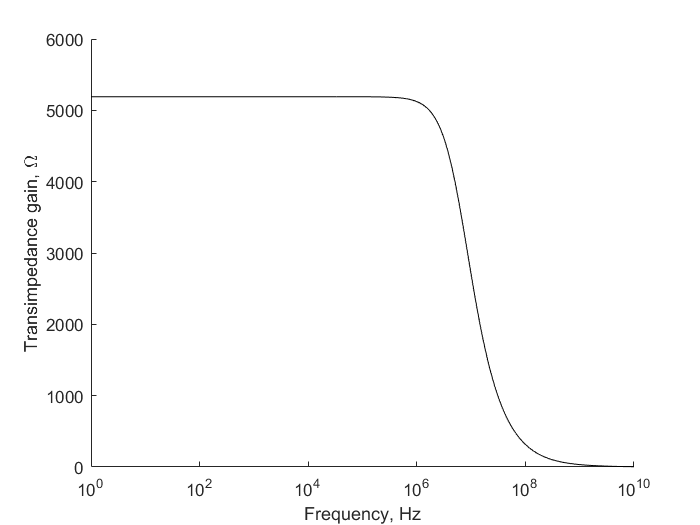}
\caption{Frequency response. Original geometric parameters, transistor only case}
\label{tranl1freq}
\end{figure}

\begin{figure}[!t]
\includegraphics[width=\columnwidth]{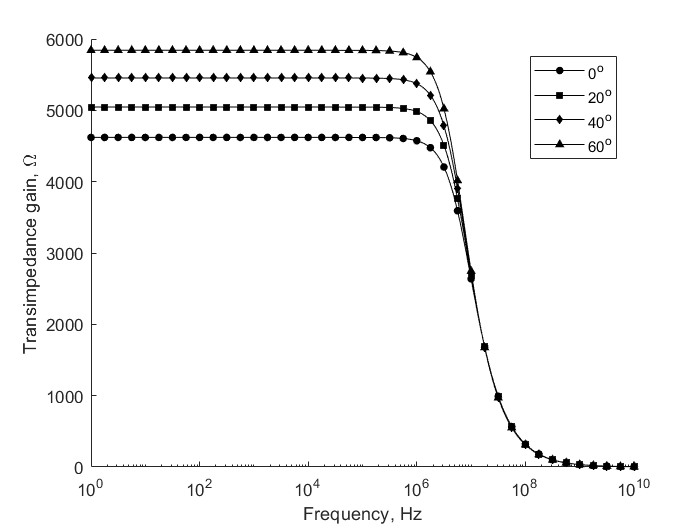}
\caption{Gain variation with temperature. Original geometric parameters, transistor only case}
\label{tranl1temp}
\end{figure}

\begin{figure}[!t]
\includegraphics[width=\columnwidth]{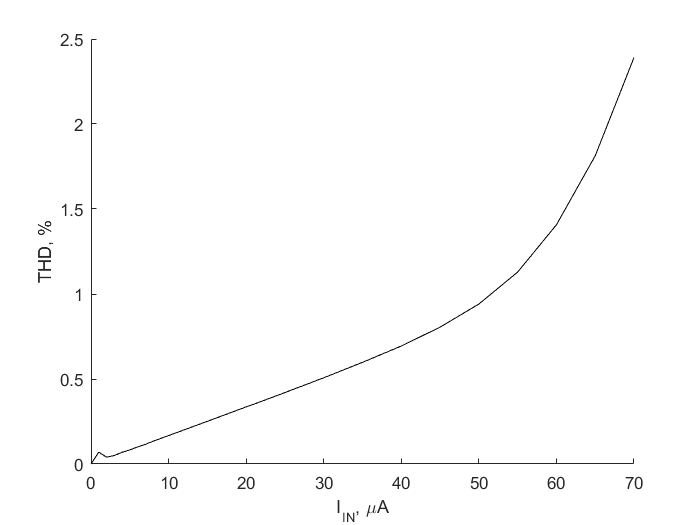}
\caption{Total Harmonic Distortion (THD) over I$_{IN}$. Original geometric parameters, transistor only case}
\label{tranl1thd}
\end{figure}

The simulations are performed using $180nm$ CMOS model for transistors and HP memristor \cite{strukov2008missing}. We simulated four different designs. In the first design, we simulate the original circuit proposed in \cite{Garcia} for $180nm$ CMOS technology with $V_{DD}=1.8V$. The sizes of the transistors are shown in Table \ref{l1sizes}. Fig. \ref{tranl1gain} shows that the range of inputs for which the amplifier has a constant gain (linear range) is from -140$\mu A$ to 60$\mu A$ for both Iin+ and Iin-.  Fig. \ref{tranl1freq} shows that the system has a 6MHz bandwidth and has a constant gain of 5.2$k\Omega$ throughout the passband. Fig. \ref{tranl1temp} shows their variation with temperature. Fig. \ref{tranl1thd} shows the total harmonic distortion for 1MHz sinusoidal input currents from 0 to 70$\mu A$. TIA has 1mV offset in output. The power dissipation is $1396\mu W$. The on-chip area of this design is $2541\mu m^2$.

\begin{table}[!t]
\caption{original transistor sizes}

\begin{tabular}{|c|c|c|c|c|c|}
\hline
 & M$_{1,2,3,4}$& M$_{5,6,7,8}$& M$_{9,10,11,12}$& M$_{13,14}$& M$_{15,16}$\\ \hline
W& 20$\mu m$& 170$\mu m$& 168.6$\mu m$& 0.707$\mu m$& 500$\mu m$\\ \hline
L& 1$\mu m$& 1$\mu m$& 1$\mu m$& 1$\mu m$& 1$\mu m$\\ \hline
\end{tabular}
\label{l1sizes}
\end{table}

\begin{figure}[!t]
\includegraphics[width=\columnwidth]{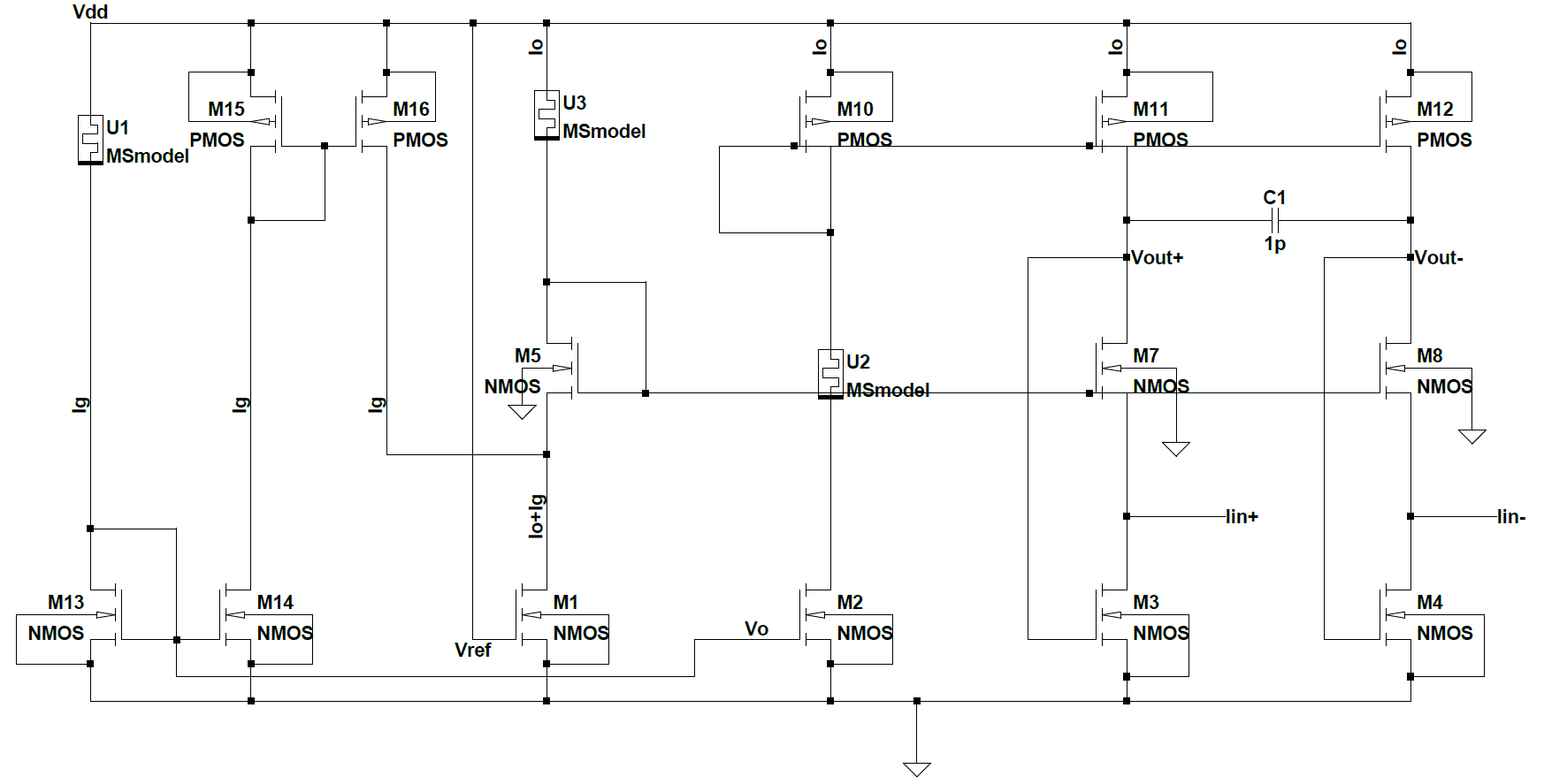}
\caption{Schematic of the differential amplifier with memristors \cite{Garcia}.}
\label{schemmem}
\end{figure}

\begin{figure}[!t]
\includegraphics[width=\columnwidth]{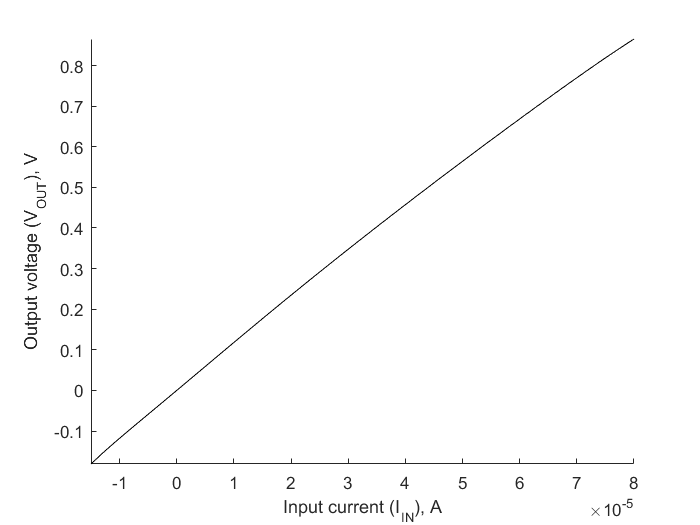}
\caption{Output voltage V$_{OUT}$ versus input current I$_{IN}$. Original geometric parameters, memristors added case}
\label{meml1gain}
\end{figure}

\begin{figure}[!t]
\includegraphics[width=\columnwidth]{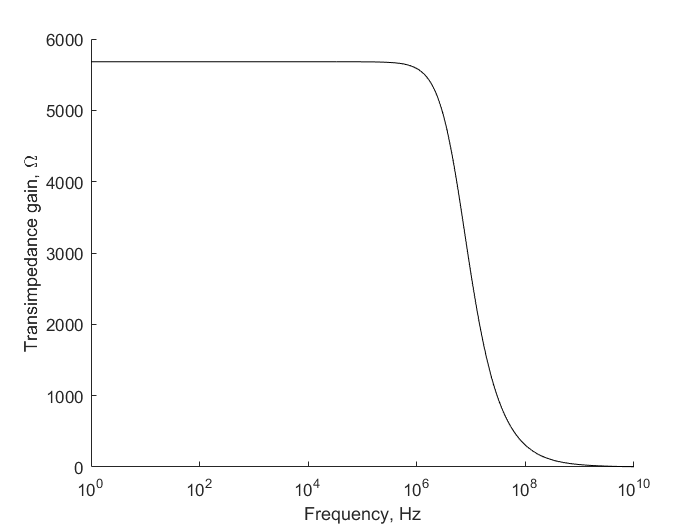}
\caption{Frequency response. Original geometric parameters, memristors added case}
\label{meml1freq}
\end{figure}

\begin{figure}[!t]
\includegraphics[width=\columnwidth]{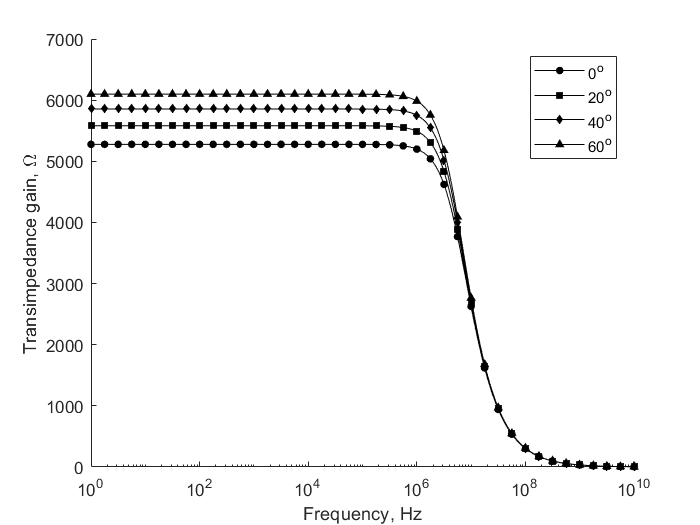}
\caption{Gain variation with temperature. Original geometric parameters, memristors added case}
\label{meml1temp}
\end{figure}

\begin{figure}[!t]
\includegraphics[width=\columnwidth]{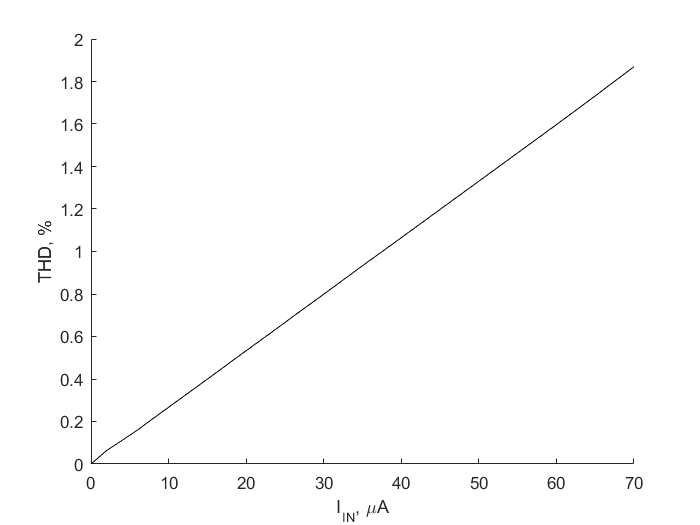}
\caption{Total Harmonic Distortion (THD) over I$_{IN}$. Original geometric parameters, memristors added case}
\label{meml1thd}
\end{figure}

In the second design, memristors U1, U2 and U3 are substituted for R1, M6 and M9, respectively. The design with memristors is shown in Fig. \ref{schemmem}. Memristors are programmed to operate in OFF state, where their resistance is equal to 10k$\Omega$. Lengths and widths of transistors were kept the same as in \cite{Garcia} and are summarized in Table \ref{l1sizes}. Fig. \ref{meml1gain} shows that the linear range of this design has been shortened and is now from -15$\mu A$ to 80$\mu A$ for both Iin+ and Iin-. Fig. \ref{meml1freq} shows that the system has a constant gain of 5.7$k\Omega$ and a 5.3MHz bandwidth. Fig. \ref{meml1temp} shows that in comparison with the first design, gain variation with temperature is considerably smaller. Fig. \ref{meml1thd} shows the total harmonic distortion with respect to input current. One can notice that this design has a linear increase THD in contrast to exponential one in the case of the first design. The output offset is 0mV. Power dissipation has been decreased to 1154$\mu W$. This design occupies 2182.4$\mu m^2$. This reduction of area was possible due to substitution of resistor R1 and transistors M6 and M9 by memristors, where the latter are assumed to have 45$nm*$90$nm$ dimensions \cite{Li}.\\

\begin{figure}[H]
\includegraphics[width=\columnwidth]{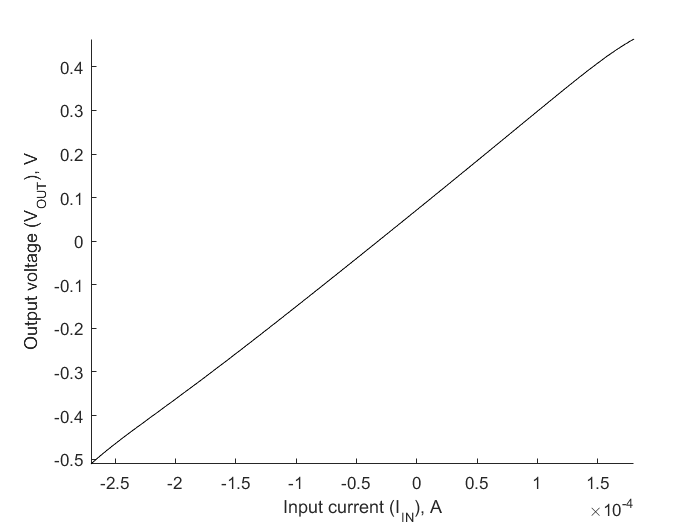}
\caption{Output voltage V$_{OUT}$ versus input current I$_{IN}$. Modified geometric parameters, transistor only case}
\label{tranl02gain}
\end{figure}

\begin{figure}[H]
\includegraphics[width=\columnwidth]{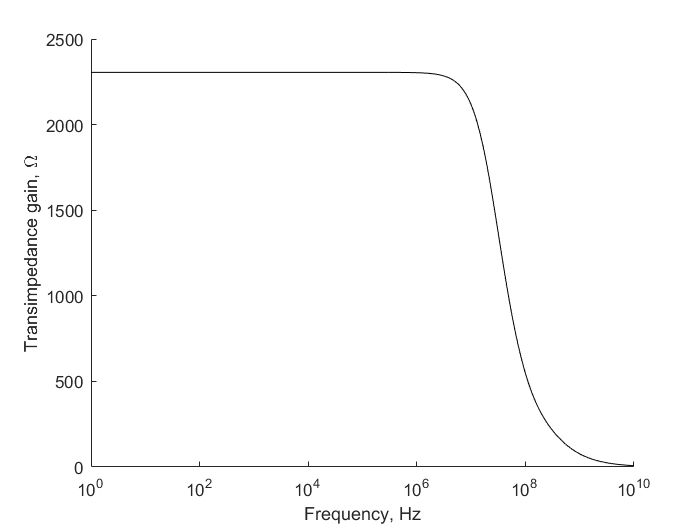}
\caption{Frequency response. Modified geometric parameters, transistor only case}
\label{tranl02freq}
\end{figure}

\begin{figure}[H]
\includegraphics[width=\columnwidth]{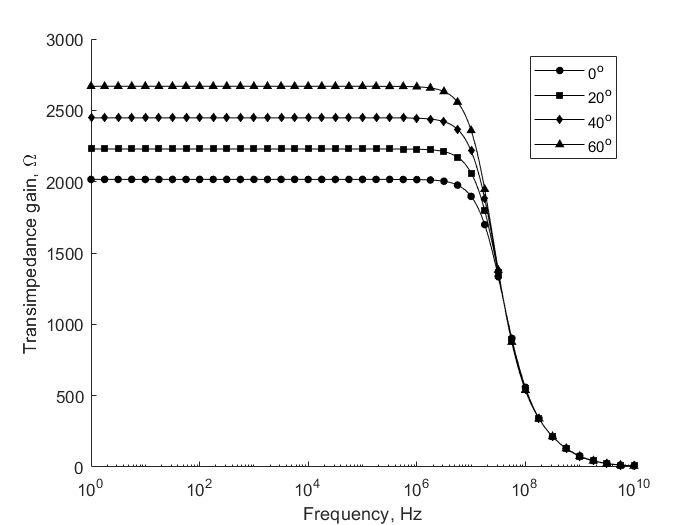}
\caption{Gain variation with temperature. Modified geometric parameters, transistor only case}
\label{tranl02temp}
\end{figure}

\begin{figure}[H]
\includegraphics[width=\columnwidth]{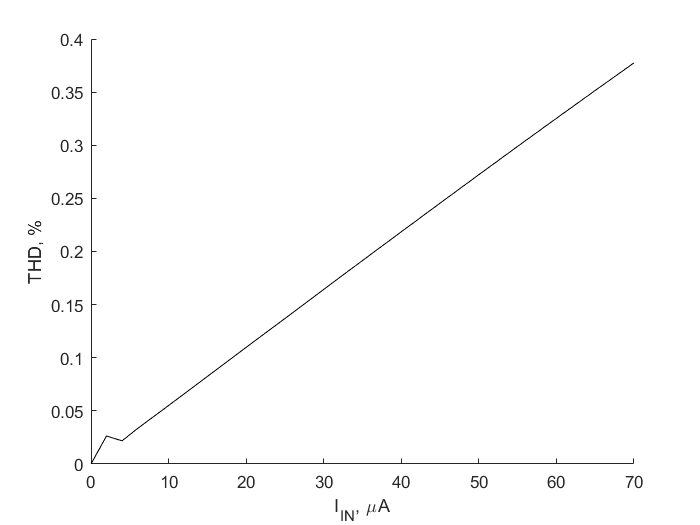}
\caption{Total Harmonic Distortion (THD) over I$_{IN}$. Modified geometric parameters, transistor only case}
\label{tranl02thd}
\end{figure}

The third design is without memristors but has length and width of each transistor decreased by the factor of 5, which is summarized in Table \ref{l02sizes}. As can be seen in Fig. \ref{tranl02gain} the linear range of this design has been increased; it is now from -270$\mu A$ to 180$\mu A$ for both inputs. Fig. \ref{tranl02freq} shows that the system has a constant gain of 2.3$k\Omega$ and a 23MHz bandwidth. Fig. \ref{tranl02temp} shows their variation with temperature. From Fig. \ref{tranl02thd}, it can be seen that THD of this design is considerably smaller than in previous designs. However, changes in geometric parameters have introduced a significant offset of 73mV in the output. Furthermore, the power dissipation has heavily grown to 2316$\mu W$. On the other hand, shortening of transistors' widths and lengths allowed a drastic decrease of the on-chip area from 2541$\mu m^2$ in the first design to 203.1$\mu m^2$.

\begin{figure}[H]
\includegraphics[width=\columnwidth]{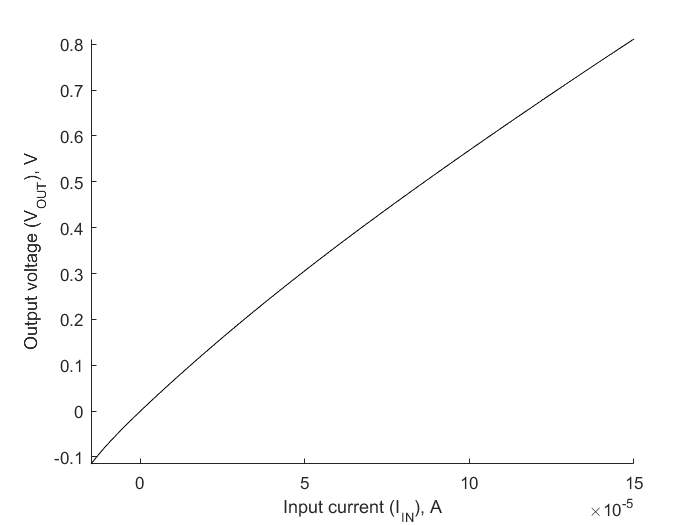}
\caption{Output voltage V$_{OUT}$ versus input current I$_{IN}$. Modified geometric parameters, memristors added case}
\label{meml02gain}
\end{figure}

\begin{figure}[H]
\includegraphics[width=\columnwidth]{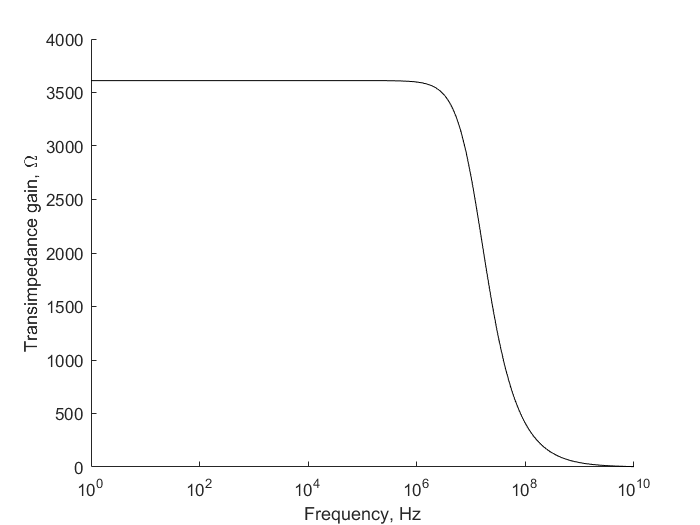}
\caption{Frequency response. Modified geometric parameters, memristors added case}
\label{meml02freq}
\end{figure}

\begin{figure}[!t]
\includegraphics[width=0.86\columnwidth]{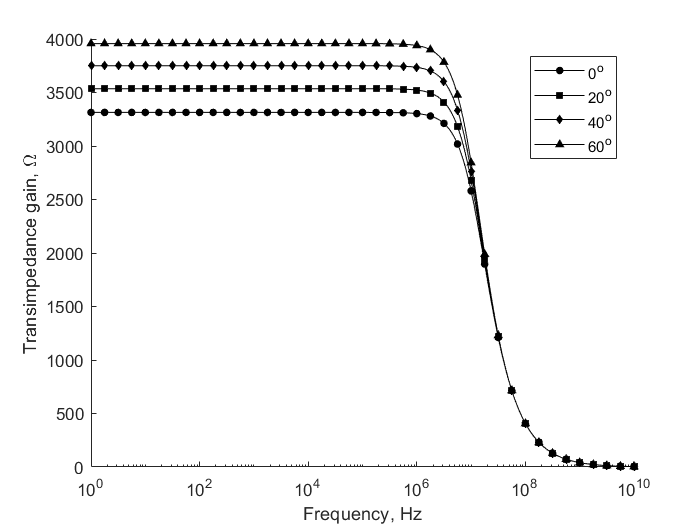}
\caption{Gain variation with temperature. Modified geometric parameters, memristors added case}
\label{meml02temp}
\end{figure}

\begin{figure}[!t]
\includegraphics[width=0.86\columnwidth]{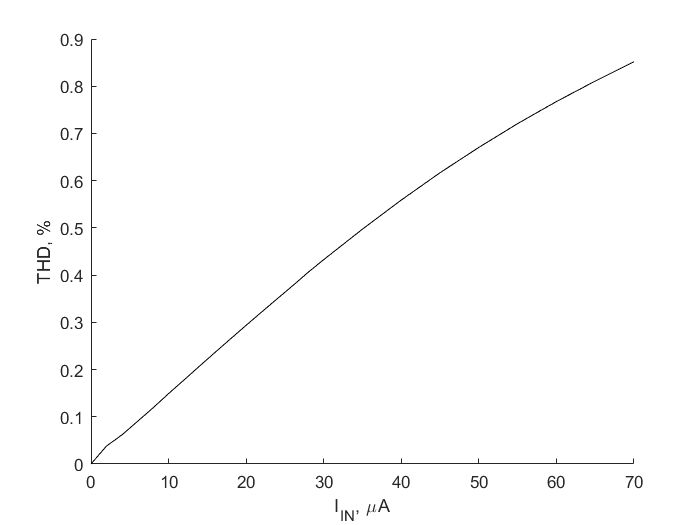}
\caption{Total Harmonic Distortion (THD) over I$_{IN}$. Modified geometric parameters, memristors added case}
\label{meml02thd}
\end{figure}

\begin{table}[!t]
\caption{Modified transistor sizes}
\centering
\begin{tabular}{|c|c|c|c|c|c|}
\hline
 & M$_{1,2,3,4}$& M$_{5,6,7,8}$& M$_{9,10,11,12}$& M$_{13,14}$& M$_{15,16}$\\ \hline
W& 4$\mu m$& 34$\mu m$& 33.72$\mu m$& 0.1414$\mu m$& 100$\mu m$\\ \hline
L& 0.2$\mu m$& 0.2$\mu m$& 0.2$\mu m$& 0.2$\mu m$& 0.2$\mu m$\\ \hline
\end{tabular}
\label{l02sizes}
\end{table}
The fourth design is the third design with memristors U1-U3 substituted instead of R1, M6 and M9. Fig. \ref{meml02gain} shows that the linear range is from -15$\mu A$ to 150$\mu A$ for both inputs. From Fig. \ref{meml02freq}, the gain is 3.6$k\Omega$ and the bandwidth is 11.3MHz. Fig. \ref{meml02temp} proofs that gain variation with temperature is smaller in designs with memristors. Fig. \ref{meml02thd} shows that total harmonic distortion of this design is higher than in the third one, but it grows at a lower pace. The substitution of memristors has completely removed the output offset. In addition, it decreased the power dissipation to 1177$\mu W$. Moreover, the on-chip area has been further reduced to 169.6$\mu m^2$\\

\section{Discussion}

Due to their small on-chip area and low power dissipation, memristors reduce the on-chip area and power dissipation. Table \ref{perf} summarizes main performance characteristics of all four designs.

Gains of all four designs significantly diverge from the theoretical value of 10$k\Omega$. Considerable impact to this mismatch is from unequal drain to source voltage drop in transistors M1-M4. It was assumed that these drops would be the same, but simulation results show that they were different. From the simulation results one can notice that application of memristors in the circuit has positive effects on gain, output offset and power dissipation. The total harmonic distortion is higher for lower input currents but it does not grow as much with increasing input as in the designs without memristors. 

However, frequency response analysis shows that bandwidth decreases when the memristors are introduced in the design. Temperature variation analysis shows that in all cases bandwidth alterations are insignificant and gain deviations are considerably smaller in the memristor based design. 

\begin{table*}[ht]
\caption{Performance characteristics}
\centering
\begin{tabular}{|c|c|c|c|c|c|}
\hline
Design& Linear range & Gain & Bandwidth & Power& Area\\ \hline
Original circuit \cite{Garcia} with $180nm$ CMOS transistors& -140$\mu A$ to 60$\mu A$& 5.2$k\Omega$& 6MHz& 1396$\mu W$& 2541$\mu m^2$\\ \hline
Memristor-based design with replaced R1, M6 and M9& -15$\mu A$ to 80$\mu A$& 5.7$k\Omega$& 5.3MHz& 1154$\mu W$&2182.4$\mu m^2$\\ \hline
Original design with decreased transistor size& -270$\mu A$ to 180$\mu A$& 2.3$k\Omega$&23MHz& 2316$\mu W$&203.1$\mu m^2$\\ \hline
Memristor-based design with decreases transistor size& -15$\mu A$ to 150$\mu A$& 3.6$k\Omega$& 11.3MHz& 1177$\mu W$& 169.6$\mu m^2$\\ \hline
\end{tabular}
\label{perf}
\end{table*}

\section{Conclusion}
Four designs which are based on \cite{Garcia} were compared and contrasted in terms of gain, frequency response, linear range, power consumption, area, total harmonic distortion and performance variations with temperature.  
 To conclude, if fully differential feature is not required, designs with memristors give many advantages over the designs without them.

\bibliographystyle{IEEEtran}
\bibliography{References.bib}

\end{document}